\pdfoutput=1
\RequirePackage{fix-cm}
\documentclass[twocolumn]{svjour3}          
\smartqed  
\usepackage{graphicx}
\usepackage[nolist]{acronym}
\usepackage{url}
\usepackage{subfigure}
\usepackage{amsmath}
\usepackage{amsfonts}
\usepackage{color}
\usepackage{algorithm}
\usepackage{algorithmic}

\begin{acronym}
	\acro{IaaS}{Infrastructure as a Service}
	\acro{VM}{virtual machine}
	\acro{JVM}{Java virtual machine}
	\acro{MMORPG}{Massively Multiplayer Online Role-Playing Game}
	\acro{MM}{directed acyclic graph}
	\acro{MTC}{many-task computing}
	\acro{HTC}{high throughput computing}
	\acro{ALT}{agglomerative likelihood tree}
	\acro{RNJ}{rooted neighbor joining}
	\acro{RTT}{round-trip time}
	\acro{DAG}{directed acyclic graph}
	\acro{QoS}{Quality of Service}
	\acro{DCSP}{Differentiated Services Code Point}
	\acro{NTP}{Network Time Protocol}
	\acro{ms}{milliseconds}
\end{acronym}

\newcommand{\labelSec}[1]{\label{sec:#1}}
\newcommand{\refSec}[1]{Section~\ref{sec:#1}}
\newcommand{\labelFig}[1]{\label{fig:#1}}
\newcommand{\refFig}[1]{Figure~\ref{fig:#1}}
\newcommand{\labelEq}[1]{\label{eq:#1}}
\newcommand{\refEq}[1]{Equation~\ref{eq:#1}}

\newcommand{\labelAlg}[1]{\label{alg:#1}}
\newcommand{\refAlg}[1]{Algorithm~\ref{alg:#1}}
\newcommand{\dw}[1]{{{\color{black}#1}}}
\newcommand{\bl}[1]{{{\color{black}#1}}}

\begin{document}

\title{Nephele Streaming: Stream Processing under QoS Constraints at Scale}



\author{
 Bj{\"o}rn Lohrmann \and
 Daniel Warneke \and
 Odej Kao
}


\institute{Bj{\"o}rn Lohrmann \at
              Technische Universit{\"a}t Berlin \\
              Einsteinufer 17\\
              10587 Berlin\\
              Germany\\
              \email{bjoern.lohrmann@tu-berlin.de}
           \and
          Daniel Warneke \at
              International Computer Science Institute (ICSI)\\
              1947 Center Street, Suite 600\\
              Berkeley, CA 94704\\
              USA\\
              \email{warneke@icsi.berkeley.edu}
           \and
          Odej Kao \at
              Technische Universit{\"a}t Berlin \\
              Einsteinufer 17\\
              10587 Berlin\\
              Germany\\
              \email{odej.kao@tu-berlin.de}
}

\date{This is a pre-print. The final publication is available at \url{link.springer.com}. It can be accessed via: \url{http://www.springer.com/alert/urltracking.do?id=L27f7914Mcc3e10Sb09c524}.} 

\maketitle

\begin{abstract}
\dw{
The ability to process large numbers of continuous data streams in a near-real-time fashion has become a crucial prerequisite for many scientific and industrial use cases in recent years. While the individual data streams are usually trivial to process, their aggregated data volumes easily exceed the scalability of traditional stream processing systems.
}

At the same time, massively-parallel data processing systems like MapReduce \dw{or Dryad currently enjoy a tremendous popularity for data-intensive applications and have proven to scale to large numbers of nodes.} Many of these systems also provide streaming capabilities. However, unlike traditional stream processors, these systems have disregarded \acs{QoS} requirements of prospective stream processing applications so far.

In this paper we address this gap. First, we analyze common design principles of today's parallel data processing frameworks and identify those principles that provide degrees of freedom in trading off the \acs{QoS} goals latency and throughput. Second, we propose a highly distributed scheme which allows these frameworks to detect violations of user-defined \dw{QoS} constraints and optimize the job execution without manual interaction. As a proof of concept, we implemented our approach for our \dw{massively-}parallel data processing framework Nephele and evaluated its effectiveness through a comparison with Hadoop Online.

For an example streaming application from the multimedia domain \bl{running on a cluster of 200 nodes, our approach improves the processing latency} by a factor of at least $13$ while preserving high data throughput when needed.

\end{abstract}

\section{Introduction} \labelSec{introduction}

In the course of the last decade, science and the IT industry have witnessed an unparalleled increase of data. While the traditional way of creating data on the Internet allowed companies to lazily crawl websites or related data sources, store the data on massive arrays of hard disks, and process it in a batch-style fashion, recent hardware developments for mobile and embedded devices together with ubiquitous networking have also drawn attention to \emph{streamed data}.

Streamed data can originate from various different sources. Every modern smartphone is equipped with a variety of sensors, capable of producing rich media streams of video, audio, and possibly GPS data. Moreover, the number of deployed sensor networks is steadily increasing, enabling innovations in several fields of life, for example energy consumption, traffic regulation, or e-health. However, \dw{an important} prerequisite to leverage those innovations is the ability to process and analyze a large number of individual data streams in a near-real-time manner. As motivation, we would like to illustrate two emerging scenarios:

\begin{itemize}
	\item \textbf{Live Media Streaming:} Today, virtually all smart phones can produce live video streams. Several websites like Justin.tv~\cite{justin.tv}, Livestream~\cite{livestream}, or Ustream~\cite{ustream} have already responded to that development \dw{and offer} their users to produce and broadcast live media content to a large audience in a way that has been reserved to major television networks before. \dw{In recent years, these platforms have been recognized to support a new form of} ``citizen journalism'' during the political incidents in the Middle East or the ``Occupy Wall Street'' movement. However, at the moment, the capabilities of those live broadcasting services are limited to media transcoding and simple picture overlays. Although the content of two different streams may overlap to a great extent (for example because the people filming the scene are standing close to each other), they are currently processed completely independent of each other. In contrast to that, future services might also offer to automatically \emph{aggregate} and \emph{relate} streams from different sources, thereby creating a more complete picture and eventually better coverage for the viewers. 

	\item \textbf{Energy \dw{I}nformatics:} Smart meters are currently being deployed \dw{in a growing number of} consumer homes by power utilities. \dw{Technically, a smart meter is a networked device that monitors} a household's power consumption and \dw{reports} it back to the power utility. On the utility's side, having such near-real-time data about power consumption is a key aspect of managing fluctuations in the power grid's load. Such fluctuations are introduced not only by consumers but also by the increasing, long-term integration of renewable energy sources. Data analytics applications that are hooked into the live meter data stream can be used for many operational aspects such as monitoring the grid infrastructure for equipment limits, initiating autonomous control actions to deal with component failures, voltage sags/spikes, and forecasting power usage. \dw{However,} especially \dw{in scenarios that involve} autonomous control actions, the freshness of the data that is being acted upon is of paramount importance. 
\end{itemize}

Opportunities to harvest the new data sources in the various domains are plentiful. However, the sheer amount of incoming data that must be processed online also raises scalability concerns with regard to existing solutions. As opposed to systems working with batch-style workloads, stream processing systems must often meet particular \ac{QoS} goals, otherwise the quality of the processing output degrades or the output becomes worthless at all. Existing stream processors \cite{borealis,aurora} have put much emphasis on meeting provided \ac{QoS} goals of applications, though often at the expense of scalability or a loss of generality~\cite{s4}.

In terms of scalability and programming generality, the predominant workhorses for data-intensive workloads at the moment are massively-parallel data processing frameworks like MapReduce~\cite{mapreduce} or Dryad~\cite{dryad}. By design, these systems scale to large numbers of nodes and are capable of efficiently transferring large amounts of data between them. Many of the newer systems~\cite{hyracks,mapreduceonline,dryad,ciel,nephelejournal} also allow to assemble complex parallel data flow graphs and to construct pipelines between the individual parts of the flow. Therefore, these systems generally are also suitable for streaming applications. However, so far they have \dw{only been used for relatively simply} streaming application, like online aggregation or ``early out'' computations~\cite{mapreduceonline}, and have not considered \ac{QoS} goals.

In this paper we attempt to bridge that gap. We have analyzed a series of open-source frameworks for parallel data processing and highlight common design principles they share to achieve scalability and high data throughput. We show how some aspects of these design principles can be used to trade off the \ac{QoS} goals latency and throughput in a fine-grained per-task manner and propose a scheme to automatically do so during the job execution based on user-defined latency constraints. Starting from the assumption that high data throughput is desired, our scheme monitors potential latency constraint violations at runtime and can then gradually \dw{apply} two techniques, \emph{adaptive output buffer sizing} and \emph{dynamic task chaining}, to met the constraints while maintaining high throughput as far as possible. As a proof of concept, we implemented the scheme for our \dw{massively-parallel} data processing framework Nephele and evaluated their effectiveness through a comparison with Hadoop Online.

This paper is an extended version of~\cite{lohrmann.2012.hpdc}. In comparison to the original work, this version extends our approach by \dw{a new, fully-distributed scheme to monitor, collect, and process the \ac{QoS} data}. Moreover, it contains the results of new experimental evaluations, conducted on a large-scale cluster \dw{with $200$ nodes} as well as an updated related work section.

The rest of this paper is structured as follows: In \refSec{dataProcessingAndStreams} we examine the common design principles of today's massively-parallel data processing frameworks and discuss the implications \dw{for} meeting the aforementioned \ac{QoS} constraints. \refSec{nepheleStream} presents our scheme to dynamically adapt to the user-defined latency constraints, whereas \refSec{evaluation} contains an experimental evaluation. \refSec{relatedWork} \dw{contrasts our work against existing stream and parallel data processors}. Finally, we conclude our paper in~\refSec{conclusion}.

\section{Massively-Parallel Data Processing and Streamed Data}\labelSec{dataProcessingAndStreams}

In recent years, a variety of frameworks for massively-parallel data analysis has emerged~\cite{hyracks,mapreduceonline,mapreduce,dryad,ciel,nephelejournal}. Many of \dw{those systems are available in an open-source version. After having} analyzed their internal structure, we found they often follow \dw{common} design principles to achieve scalability and high throughput.

\dw{In this section we highlight} those \dw{common design} principals and discuss their implications on stream processing under \ac{QoS} constraints.

\subsection{Design Principles of \dw{Massively-}Parallel Data Processing Frameworks}\labelSec{principalFrameworkProperties}

\begin{figure*}[t]
	\centering
		\includegraphics[width=15cm]{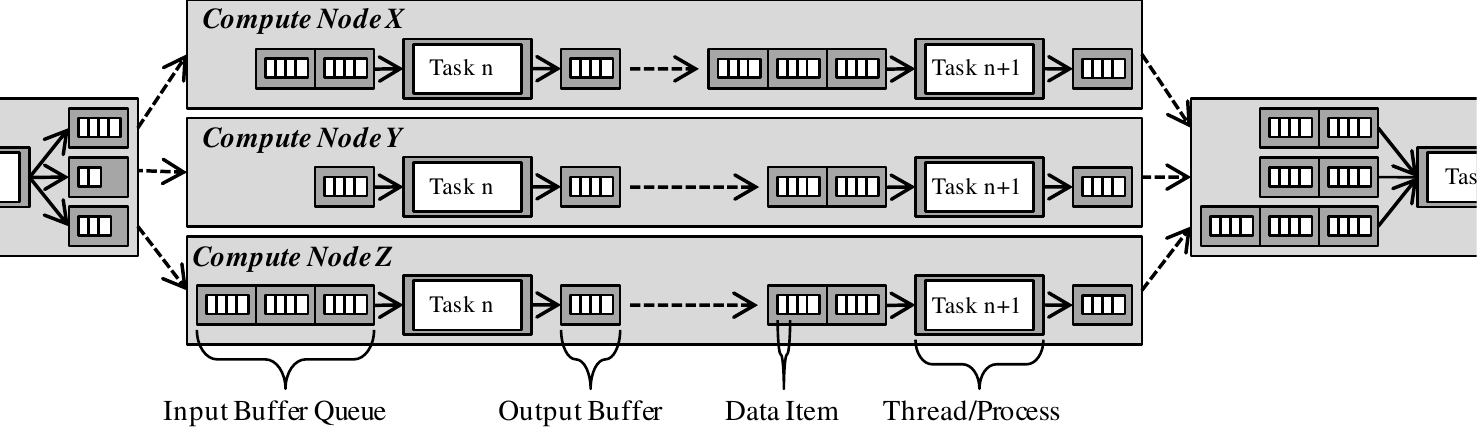}
		\caption{Typical processing pattern of frameworks for massively-parallel data analysis}
	\labelFig{streamModel}
\end{figure*}

Frameworks for \dw{massively}-parallel data processing typically follow a master-worker pattern. The master node receives jobs from the user, splits them into sets of individual tasks, and schedules those tasks to \dw{be executed} on the available worker nodes.

The structure of those jobs can usually be described by a graph with vertices representing the job's individual tasks and the edges denoting communication channels between them. For example, from a high-level perspective, the graph representation of a typical MapReduce job would consist of a set of Map vertices connected to a set of Reduce vertices. Some frameworks have generalized the MapReduce model to arbitrary \acp{DAG}~\cite{hyracks,dryad,nephelejournal}, some even allow graph structures containing loops~\cite{ciel}.

However, independent of the concrete graph model used to describe \dw{the} jobs for the respective \dw{parallel processing} framework, the way both the vertices and edges translate to system resources at runtime is surprisingly similar among all of these systems.

Each task vertex of the overall job typically translates to either a separate process or a separate thread at runtime. Considering the large number of CPUs (or CPU cores) these frameworks must scale up to, this is a reasonable design decision. By assigning each task to a different thread/process, those tasks can be executed independently and utilize a separate CPU core. Moreover, it gives the underlying operating system various degrees of freedom in scheduling the tasks among the individual CPU cores. For example, if a task cannot fully utilize its assigned CPU resources or is waiting for an I/O operation to complete, the operating system can assign the idle CPU time to a different thread/process.

\dw{In most cases, the} communication model of parallel data processing systems follows a producer-consumer pattern. Tasks can produce a sequence of \textit{data items} which are then passed to and consumed by their successor tasks according to the edges of the job's graph representation. The way the data items are physically transported from one task to the other depends on the concrete framework. In the most lightweight case, two tasks are represented as two different threads running inside the same operating system process and can use shared memory to exchange data. If tasks are mapped to different processes, possibly running on different worker nodes, the data items are typically exchanged through files or a network connection.

However, since all of these frameworks have been designed for data-intensive workloads and hence strive for high data throughput, they attempt to minimize the transfer overhead per data item. As a result, these frameworks try to avoid shipping individual data items from one task to the other. As illustrated in \refFig{streamModel}, the data items produced by a task are typically collected in a larger \textit{output buffer}. Once its capacity limit has been reached, the entire buffer is shipped to the receiving task and in many cases placed in its \emph{input buffer queue}, waiting to be consumed.

\subsection{Implications for \ac{QoS}-Constrained Streaming Applications}

\begin{figure*}[t]
	\centering
	\mbox{
		\subfigure[Average data item latency]{
			\includegraphics[width=8.4cm]{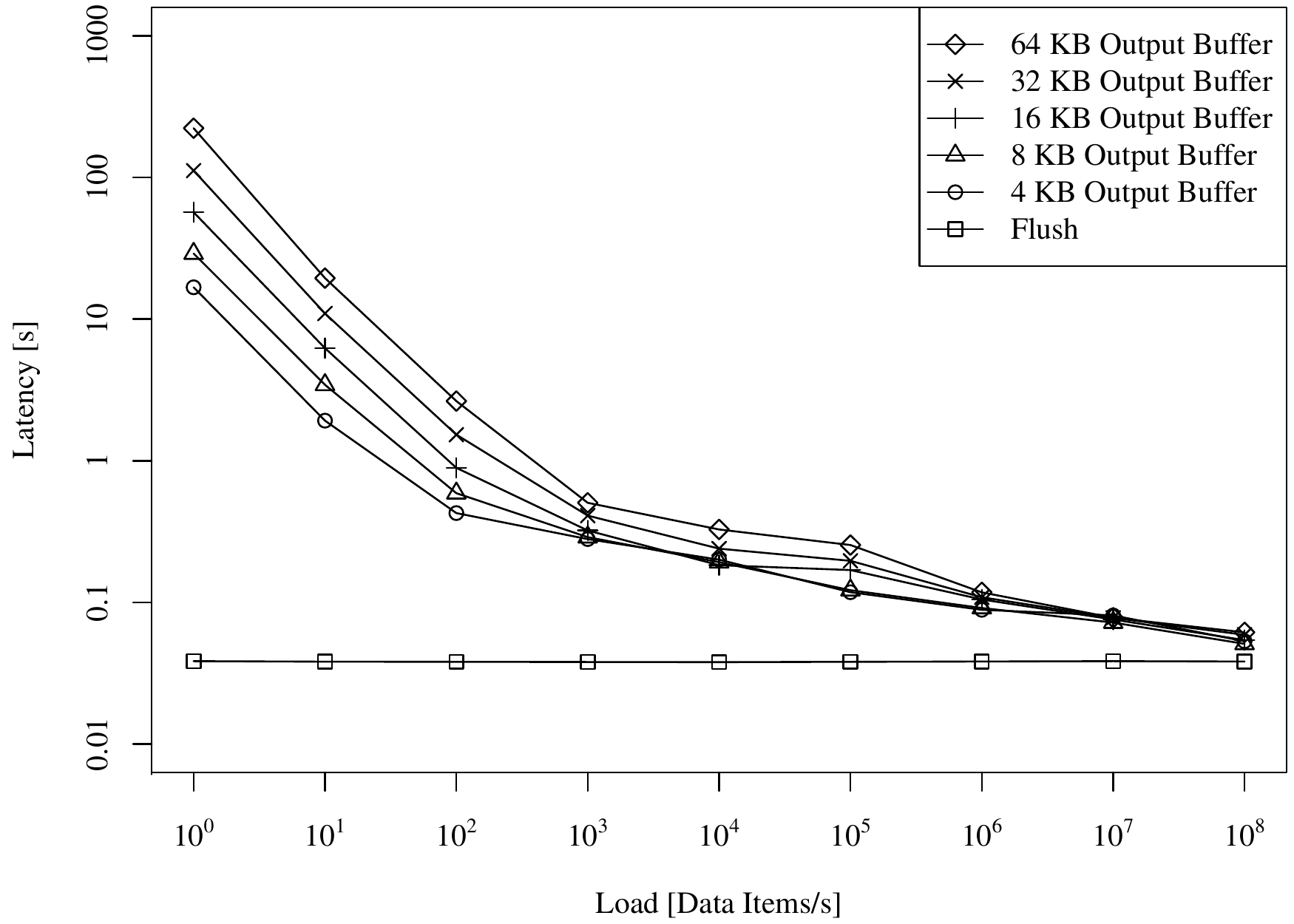}
			\labelFig{qosTradeoffLatency}
		}

		\subfigure[Average data item throughput]{
			\includegraphics[width=8.4cm]{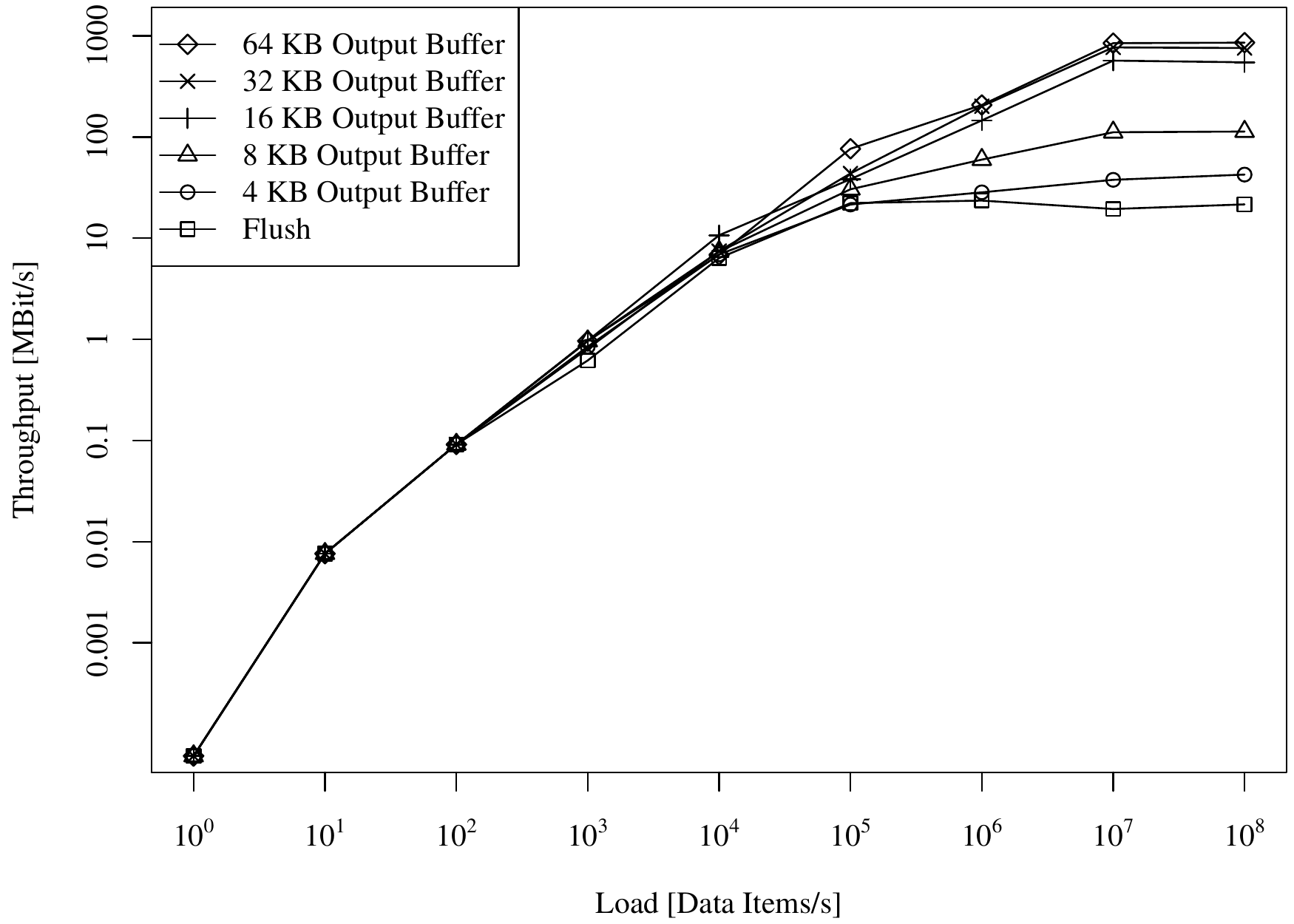}
			\labelFig{qosTradeoffThroughput}
		}
	}
	
		\caption{The effect of different output buffer sizes on data item latency and throughput}
	\labelFig{qosTradeoff}
\end{figure*}

\dw{After having} highlighted some basic design principles of today's \dw{massively-}parallel data processing frameworks, we now discuss which aspects of those principles provide degrees of freedom in trading off the different \ac{QoS} goals latency and throughput.

\subsubsection{The Role of the Output Buffer}

As explained previously, most frameworks for parallel data processing introduce distinct output buffers to minimize the transfer overhead per data item and improve the data item throughput, i.e.\ the average number of items that can be shipped from one task to the other in a given time interval.

For the vast majority of data processing frameworks we have analyzed in the scope of our research, the output buffer size could be set on a system level, i.e.\ all jobs of the respective framework instance were forced to use the same output buffer sizes. Some frameworks also allowed to set the output buffer size per job, for example Apache Hadoop~\cite{hadoop}. Typical sizes of these output buffers range from several MB \dw{down} to \dw{$8$ or $4$} KB, depending on the focus of the framework.

While \dw{output buffers are pivotal to achieve high data throughputs}, they also make it hard to optimize jobs for current \dw{massively-}parallel data processors towards the \ac{QoS} goal latency. Since an output buffer is typically not shipped until it has reached its capacity limit, the latency \dw{which} an individual data item experiences depends on the system load.

To illustrate this effect, we created a small sample job consisting of two tasks, a sender task and a receiver task. The sender created data items of $128$ bytes length at a fixed rate $n$ and wrote them to an output buffer of a fixed size. Once an output buffer had reached its capacity limit, it was sent to the receiver through a TCP connection. We ran the job several times. \dw{After} each run, we varied the output buffer size.

The results of this initial experiment are depicted in~\refFig{qosTradeoff}. As illustrated in~\refFig{qosTradeoffLatency}, the average latency from the creation of a data item at the sender until its arrival at the receiver depends heavily on the creation rate and the size of the output buffer. With only one created data item per second and an output buffer size of $64$ KB, it takes more than $222$ seconds on an average before an item arrives at the receiver. At low data creation rates, the size of the output buffer has a significant effect on the latency. The more the data creation rate increases, the more the latency converges towards a lower bound. At a rate of $10^8$ created items per second, we measured an average data item latency of approximately $50$ \ac{ms}, independent of the output buffer size.

As a baseline experiment, we also executed separate runs of the sample job which involved flushing incomplete output buffers. Flushing forced the system to transfer the output buffer to the receiver after each written data item. As a result, the average data item latency was uniformly $38$ \ac{ms}, independent of the data creation rate.

\refFig{qosTradeoffThroughput} shows the effects of the different data creation rates and output buffer sizes on the throughput of the sample job. While the \ac{QoS} objective latency suggests using small output buffers or even flushing incomplete buffers, these actions show a detrimental effect when high data throughput is desired. As depicted in~\refFig{qosTradeoffThroughput}, the data item throughput that could be achieved grew with the size of the output buffer. With relatively \dw{large} output buffers of $64$ or $32$ KB in size, we were able to fully saturate the $1$ GBit/s network link between the sender and the receiver, given a sufficiently high data creation rate. However, the small output buffers failed to achieve a reasonable data item throughput. In the most extreme case, i.e.\ flushing the output buffer after every written data item, we were unable to attain a data item throughput of more than $10$ MBit/s. The reason for this is the disproportionately high transfer overhead per data item (output buffer meta data, memory management, thread synchronization) that parallel data processing frameworks in general are not designed for. \dw{A} similar behavior is known from the TCP networking layer, where the Nagle algorithm can be deactivated (TCP\_NODELAY option) to improve connection latency.

In sum, the sample job highlights an interesting trade-off that exists in current data processing frameworks with respect to the output buffer size. While jobs with low latency demands benefit from small output buffers, the classic data-intensive workloads still require relatively large output buffers in order to achieve high data throughput. This trade-off puts the user in charge of configuring a reasonable output buffer size for his job and assumes that (a) the used processing framework allows him to specify the output buffer size on a per-job basis, (b) he can estimate the expected load his job will experience, and (c) the expected load does not change over time. In practice, however, at least one of those three assumptions often does not hold.  One might also argue that there is no single reasonable output buffer size for an entire job as the job consists of different tasks that produce varying data item sizes at varying rates, so that any chosen fixed output buffer size can only result in acceptable latencies for a fraction of the tasks but not for all of them. 

\subsubsection{The Role of the Thread/Process Model}

Current frameworks for parallel data processing typically map different tasks to different operating system processes or at least different threads. While this facilitates natural scalability and load balancing between different CPUs or CPU cores, it also raises the communication overhead between tasks. In the most lightweight case, where different tasks are mapped to different threads within the same process and communication is performed via shared memory, the communication overhead typically only consists of thread synchronization, scheduling, and managing cache consistency issues. However, when the communicating tasks are mapped to different processes or even worker nodes, passing data items between them additionally involves serialization/deserialization and, depending on the way the data is  exchanged, writing the serialized data to the network/file system and reading it back again.

Depending on the complexity of the tasks, the communication overhead can account for a significant fraction of the overall processing time. If the tasks themselves are lightweight, but the data items are rather large and complex to serialize/deserialize (as in case of a filter operation on a nested XML structure~\cite{alexandrov.2011.btw}), the overhead can limit the throughput and impose a considerable processing latency.

\begin{figure}[th]
	\centering
		\subfigure[Pipeline without task chaining]{
		\includegraphics[width=7.2cm]{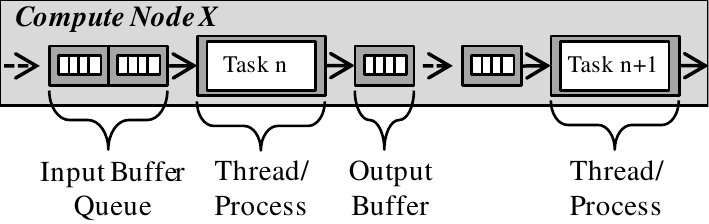}
		\labelFig{executionWithoutChaining}
	}

	\subfigure[Pipeline with task chaining]{
		\includegraphics[width=7.2cm]{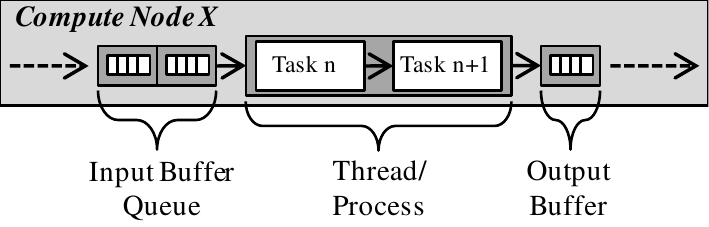}
		\labelFig{executionWithChaining}
	}
	
		\caption{Different execution models with and without task chaining}
	\labelFig{chaining}
\end{figure}

As illustrated in~\refFig{chaining}, a common approach to address this form of communication overhead is to chain lightweight tasks together and execute them in a single thread/process. The most popular example in the area of parallel data processing is probably the chained map functions from Apache Hadoop. However, a similar idea was also described earlier as rewriting a program to its ``normal form'' by Aldinucci and Danelutto~\cite{aldinucci.1999.pdcs} in the context of stream parallel skeletons.

Before starting a Hadoop job, a user can specify a series of map functions to be chained. Hadoop will then execute these functions in a single process. Chaining tasks often also eliminates the need for separate output buffers. For example, in case of Hadoop's chained map functions, the user code of the next map function in the processing chain can be directly invoked on the previous map function's output. Depending on the semantics of the concatenated tasks, chaining may also render the serialization/deserialization between tasks superfluous. If the chained tasks are stateless (as typically expected from map functions in Hadoop), it is safe to pass the data items from one task to the other by reference.

With regard to stream processing, chaining tasks \dw{is an} interesting approach to reduce processing latency and increase throughput at the same time. However, similar to the output buffer size, there might also be an important trade-off, \dw{especially when the job's workload is unknown in advance or likely to change over time}.

\dw{Currently}, task chaining is performed at compile time, so once the job is running, all chained tasks are bound to a single execution thread. In situations with low load, this might be beneficial since communication overhead is decreased and potential throughput and latency goals can be met more easily. However, when the load increases in the course of the job processing, the static chaining prevents the underlying operating system from distributing the tasks across several CPU cores. As a result, task chaining can also be disadvantageous if (a) the complexity of the chained tasks is unknown in advance or (b) the workload the streaming job has to handle is unknown or changes over time.

\section{Automated QoS-Optimization for Streaming Applications}\labelSec{nepheleStream}

Currently, it is the user of a particular \dw{parallel data processing} framework who must estimate the effects of the configured \dw{output} buffer size and thread/process model on a job's latency and throughput characteristics in a cumbersome and inaccurate manner.

In this section, we propose an extension to parallel data processing frameworks which spares the user this hassle. Starting from the assumption that high throughput continues to be the predominant \ac{QoS} goal in parallel data processing, our extension lets users add latency constraints to their job specifications. Based on these constraints, it continuously monitors the job execution and detects violations of the provided latency constraints \emph{at runtime}. Our extension can then selectively trade high data throughput for a lower processing latency using two distinct strategies, \emph{adaptive output buffer sizing} and \emph{dynamic task chaining}.

As a proof of concept, we implemented this extension as part of our massively-parallel data processing framework Nephele~\cite{nephelejournal},  which runs data analysis jobs based on \acp{DAG}. However, based on the common principles identified in the previous section, we argue that similar strategies are applicable to other \dw{parallel data processing} frameworks as well.

\begin{figure}
	\centering
		\includegraphics[width=8.4cm]{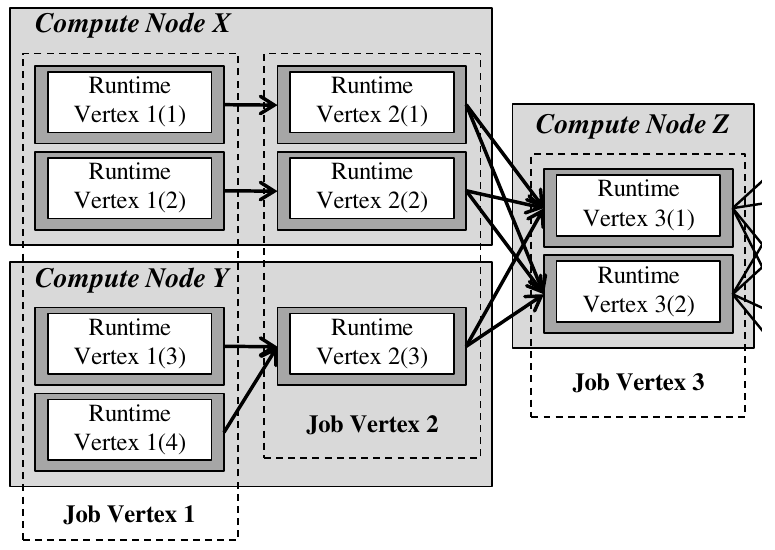}
		\caption{Exemplary job graph with embedded runtime graph and worker node allocation}
	\labelFig{graphEmbedding}
\end{figure}

\subsection{Prerequisites}
For the remainder of the paper, we will assume that the underlying structure of a job is a \ac{DAG}. For a given job we will formally distinguish between two representations, the \textit{job graph} and the \textit{runtime graph}. \bl{The job graph is a compact description of the of the job's structure provided by the user and serves as a template for constructing the runtime graph. The runtime graph is then derived from the job graph by the execution framework once the job is started.

\subsubsection{The Job Graph}
The job graph is provided by the user and indicates to the framework which user code to run and with which degree of parallelism this should be done. It shall be defined as a \ac{DAG} $JG=(JV,JE)$ that consists of \textit{job vertices} $jv \in JV$ connected by directed \textit{job edges} $je \in JE$.

\subsubsection{The Runtime Graph}
The runtime graph is a parallelized version of the job graph to be used by the execution framework during job execution. It shall be defined as a DAG $G=(V, E)$ where each \textit{runtime vertex} $v \in V$ is a task containing user code. The directed \textit{runtime edge} $e=(v_1, v_2)\in E$ is a channel along which the task $v_1$ can send data items of arbitrary size to task $v_2$. The terms task and runtime vertex, as well as channel and runtime edge will be used synonymously.

At runtime, each job vertex is equivalent to a set of vertices in the runtime graph. For notational simplicity we shall sometimes view job vertices as sets $jv \subseteq V$. Analogous, each job edge $je \subseteq E$ shall be regarded as a set of runtime graph edges. 

Each vertex of the runtime graph is allocated to a \textit{worker node} by the framework. We will denote this mapping by $worker(v) \in W$, where $W$ is the set of all worker nodes.}

\refFig{graphEmbedding} shows the underlying structure of a job in an aggregated form, depicting both the job graph as well as the runtime graph. Note that the details of how the mapping between job and runtime graph is constructed is left up to the framework and that we do not make any assumption other than the existence of the two graphs and the existence of a relationship.

\subsection{Specifying Latency Constraints}
In order to specify latency constraints, a user must be aware how much latency his application can tolerate in order to still be useful. With his knowledge from the application domain a user should then identify latency critical series of vertices and edges within the job graph for which he can express required upper latency bounds in the form of \textit{job constraints}. Since the job constraints are part of the job description provided by the user, they must be attached to the job graph and indicate to the framework which portions of the runtime graph to monitor and optimize.

In the following, we will first introduce the runtime-level notions of task, channel, and sequence latency. Based on these, we will define the semantics of user-provided latency constraints on the job graph.

\subsubsection{Task Latency}\labelSec{taskLatency}
Given three tasks $v_i, v_x, v_y \in V$, an incoming channel $e_{in}=(v_x,v_i)$ and an outgoing channel $e_{out}=(v_i,v_y)$, we shall define the \textit{task latency} $tl(d, v_i, v_x{\rightarrow}v_y)$ as the time difference between a data item $d$ entering the user code of $v_i$ via the channel $e_{in}$ and the next data item exiting the user code via $e_{out}$.

This definition has several implications. First, task latency is undefined on source and sink tasks as these task types lack incoming \dw{or}, respectively, outgoing channels. Task latencies can be infinite if the task never emits for certain in/out channel combinations. Moreover, task latency can vary significantly between subsequent items, for example, if the task reads two items but emits only one item after it has read the last one of the two. In this case the first item will have experienced a higher task latency than the second one.

\subsubsection{Channel Latency} Given two tasks $v_i, v_j \in V$ connected via channel $e=(v_i,v_j) \in E$, we define the \textit{channel latency} $cl(d, e)$ as the time difference between the data item $d$ exiting the user code of $v_i$ and entering the user code of $v_j$. The channel latency may also vary significantly between data items on the same channel due to differences in item size, output buffer utilization, network congestion, and queues that need to be transited on the way to the receiving task.

\subsubsection{Sequence Latency}\labelSec{sequenceLatency}
We shall define a sequence as \bl{an n-tuple} of connected tasks and channels. Sequences can thus be used to identify the parts of the runtime graph for which the application has latency requirements.

Let us assume a sequence $S=(s_1, \ldots, s_n)$, $n \geq 1$ of connected tasks and channels. The first element of the sequence is allowed to be either a task or a channel, the same holds for the last element. For example, if $s_2$ is a task, then $s_{1}$ needs to be an incoming and $s_{3}$ an outgoing channel of the task. If a data item $d$ enters the sequence $S$, we can define the \textit{sequence latency} $sl(d,S)$ that the item $d$ experiences as $sl^*(d,S,1)$ where

\begin{equation*}
sl^*(d, S, i)=
	\begin{cases}
	l(d, s_i) + sl^*(s_i(d),S,i+1) & \text{if } i<n\\
  l(d, s_i) & \text{if } i=n\\
	\end{cases}
\end{equation*}

If $s_i$ is a task, then $l(d,s_i)$ is equal to the task latency $tl(d, s_i, v_x{\rightarrow}v_y)$ and $s_i(d)$ is the next data item emitted by $s_i$ to be shipped via the channel $(s_i,v_y)$. If $s_i$ is a channel, then $l(d, s_i)$ is the channel latency $cl(d,s_i)$ and $s_i(d)=d$.
  
\subsubsection{Latency Constraints}
\bl{
When the user has identified latency critical portions of the job graph, he can express his requirements as latency constraints on the respective parts of the job graph. Similar to the way the runtime graph is derived from the job graph, a framework can derive \textit{runtime latency constraints} from user-provided \textit{job latency constraints}. We will first introduce a formal notion of job latency constraints and then describe how the relationship between job and runtime graph can be used to derive runtime latency constraints.

\paragraph{Job Latency Constraints}
Analogous to the runtime-level sequence introduced in \refSec{sequenceLatency} we can define a job-level sequence. A \textit{job sequence} $JS$ shall be defined as an n-tuple of connected vertices and edges within the job graph, where both the first and last element can be a job vertex or a job edge. Each $JS$ is hence equivalent to a set of sequences $\{S_1, \ldots, S_n\}$ within the runtime graph.

For latency critical job sequences, the user can express his or her maximum tolerable latency as a set of \textit{job constraints} $JC=\{jc_1, \ldots, jc_n\}$ to be attached to the job graph. Each such constraint $jc_i=(JS, l, t)$ expresses a desired upper latency limit $l$ for the data items passing through all the runtime-graph sequences of $JS_i$  during any time span of $t$ time units.

\paragraph{Runtime Latency Constraints} A given job constraint $jc=(JS, l, t)$ induces a set of \textit{runtime constraints} $C=\{C_1,\ldots,C_n\}$. Each runtime constraint $C=(S_i, l, t)$ is induced by exactly one of the runtime sequences of $JS$. Such a runtime constraint expresses a desired upper latency limit $l$ for the arithmetic mean of the sequence latency $sl(d,S_i)$ over all the data items $d\in D_t$ that enter the sequence $S_i$ during any time span of $t$ time units:
\begin{equation}\labelEq{constraint}
 \frac{\sum_{d\in D_t}sl(d,S_i)}{|D_t|} \leq l_{S_i}
\end{equation}
}

Note that a runtime constraint does not specify a hard upper latency bound for every single data item but only a ``statistical'' upper bound over the items running through the workflow during the given time span $t$. While hard upper bounds for each item may be desirable, we doubt that meaningful hard upper bounds can be achieved considering the complexity of most real-world setups in which such parallel data processing frameworks are deployed. \bl{In this context the purpose of the time span $t$ is to provide a concrete time frame for which the violations of the constraint can be tested. With $t \rightarrow \infty$ the constraint would cover all data items ever to pass through the sequence of tasks and channels. In this case, it is not possible to evaluate during the job's execution whether or not the constraint has been violated as we may be dealing with a possibly infinite stream of items.}

\subsection{Measuring Workflow Latency}\labelSec{measuring}
In order to make informed decisions where to apply optimizations to a running workflow we designed and implemented means of sampling and estimating the latency of a sequence. The master node that has global knowledge about the defined latency constraints will instruct the worker nodes about where they have to perform latency measurements. For the elements (task or channel) of each constrained sequence, latencies will be measured on the respective worker node once by during a configured time interval, the \textit{measurement interval}. This scheme can quickly produce high numbers of measurements with rising numbers of tasks and channels. \bl{For this reason, each node runs a \textit{QoS Reporter} that locally preaggregates measurement data on the worker node and prepares a report for each \textit{QoS Manager} it has to report to. For which QoS Managers reports must be sent is determined by the scheme described in \refSec{locatingConstraintViolations}.} To avoid bursts of reports, the QoS Reporter chooses a random offset for the reports of each QoS Manager. Each report contains the following data:
\begin{enumerate}
 \item An estimation of the average \textit{channel latency} of the locally incoming channels (i.e. it is an incoming channel on the worker node) of the constrained sequences that the QoS Manager is interested in. The average latency of a channel is estimated using \textit{tagged} data items. A tag is a small piece of data that contains a creation timestamp and a channel identifier and it is added when a data item exits the user code of the channel's sender task and is evaluated just before the data item enters the user code of the channel's receiver task. The QoS Reporter on the receiving worker node will then add the measured latency to its aggregated measurement data. The tagging frequency is chosen in such a way that we have one tagged data item during each measurement interval if there \dw{is} any data flowing through the channel. If the sending and receiving tasks are executed on different worker nodes, clock synchronization is required. 

 \item The average \textit{output buffer lifetime} for each locally outgoing channel of the constrained sequences that the QoS Manager is interested in. This is the average time it took for output buffers to be filled.

 \item An estimation of the average \textit{task latency} for each task of the constrained sequences that the QoS Manager is interested in. Task latencies are measured in an analogous way to channels, but here we do not require tags. Once every measurement interval, a task will note the difference in system time between a data item entering the user code and the next data item leaving it on the channels specified in the constrained sequences. Again, the measurement frequency is chosen in a way that we have one latency measurement during each measurement interval.
\end{enumerate}

As an example, let us assume a constrained sequence $S=(e_1,v_1,e_2)$. Tags will be added to the data items entering channel $e_1$ once every measurement interval. Just before a tagged data item enters the user code of $v_1$, the tag is removed from the data item and the difference between the tag's timestamp and the current system time is added to the locally aggregated measurement data. Let us assume a latency measurement is required for the task $v_1$ as well. In this case, just before handing the data item to the task, the current system time is stored in the task's environment. The next time the task outputs a data item to be shipped via channel $e_2$ the difference between the current system time and the stored timestamp is again added to the locally aggregated measurement data. Before handing the produced data item to the channel $e_2$, the worker node may choose to tag it, depending on whether we still need a latency measurement for this channel. Once every measurement interval the QoS Reporters on the worker nodes flush their reports with the aggregated measurement data to the assigned QoS Managers.

A QoS Manager stores the reports it receives from its reporters. For a given constraint $(S_i, l_{S_i}, t) \in C$, it will keep all latency measurement data concerning the elements of $S_i$ that are fresher than $t$ time units and discard all older measurement data. Then, for each element of $S_i$, it will compute a running average over the measurement values and add the results up to an estimation of the left side of \refEq{constraint}. The accuracy of this estimation depends mainly on the chosen measurement interval.

The aforementioned \textit{output buffer lifetime} measurements are subjected to the same running average procedure. To the running average of the \textit{output buffer lifetime} of channel $e$ over the past $t$ time units we shall refer as $oblt(e,t)$.  Note that the time individual data items spend in output buffers is already contained in the channel latencies, hence we do not need the output buffer lifetime to estimate sequence latencies. It does however play the role of an indicator when trying to locate channels where the output buffer sizes can be optimized (see \refSec{constraintViolationReaction}).

\subsection{Locating Constraints Violations}\labelSec{locatingConstraintViolations}
The task of analyzing all of the measurement data and locating latency constraints can quickly overwhelm any central node.  While it may still be possible for a central node to keep all of the measurement data in memory, it is impractical to repeatedly search through the set $C$ of all runtime constraints in order to detect constraint violations. For large runtime graphs, even explicitly materializing all runtime constraints can be infeasible. As an example, consider a DAG such as the one in \refFig{evaluationJob}. Due to the amount of channels between the \textit{Partitioner} and \textit{Decoder}, as well as between the \textit{Encoder} and \textit{RTP Server} tasks, the number of sequences with latency constraints grows quickly with the degree of parallelism. For this specific graph, the number of constrained runtime sequences is $m^3$, where $m$ is the degree of parallelism between tasks of the same type, hence for $m=800$ we obtain $512\times10^6$ constrained sequences. Therefore, we chose to distribute the work of locating and reacting to constraint violations in order to minimize the impact on a running job.

In the following we will first provide an overview of our distributed \ac{QoS} management scheme and then provide details on how such a structure can be set up for a framework following \dw{a} master-worker pattern.

\subsubsection{Distributed QoS Management Overview}
When \dw{the} master node receives the job description with attached latency constraints from a user, it schedules the tasks as usual to run on the available worker nodes. However, besides executing the scheduled tasks, worker nodes are also responsible for independently monitoring constraints and reacting to constraints violations. For this purpose, the master node assigns the roles of \textit{QoS Reporter} and \textit{QoS Manager} to selected worker nodes.

\paragraph{QoS Reporter Role}
A worker node with this role runs a background process that collects measurement data for all of the tasks and channels which are local to the worker node and part of a constrained runtime sequence. It collects the measurement data described in \refSec{measuring} and also knows which measurement values to send to which QoS Manager. Reports that aggregate measurement data for the QoS Managers are sent once every measurement interval on an as-needed basis, i.e. no empty reports are sent.

\paragraph{QoS Manager Role}
A worker node with this role runs a background process that analyzes the measurement data it receives from its QoS Reporters. For this purpose, the QoS Manager is equipped with a subgraph of the original runtime graph. This subgraph both stores the measurement data and can be used to efficiently enumerate violated runtime constraints. Upon detection of a constraint violation a QoS Manager can initiate countermeasures to improve latency as described in \refSec{constraintViolationReaction}. 

\subsubsection{Distributed QoS Management Setup}
For large \ac{DAG}s the main complexity lies in assigning the QoS Manager role to the available worker nodes. We will briefly discuss our objectives when designing our approach to QoS Manager Setup and then propose an algorithm to efficiently allocate the QoS Manager role even for large runtime graphs.

\paragraph{Objectives}
\bl{The main objective is to split the runtime graph $G$ into $m$ subgraphs $G_i=(V_i,E_i)$ each of which is to be assigned to a QoS Manager while meeting the following conditions:
\begin{enumerate}
 \item The number $m$ of subgraphs is maximized. This ensures that the amount of work to be done by each QoS Manager is minimized and thus reduces the impact on the job.
 \item The number of common vertices between subgraphs should be minimized:
 \begin{equation*}
\underset{G_1, \ldots, G_m}{\text{minimize}} \sum_{0 \leq i < m}\sum_{j\neq i}|V_i \cap V_j|
\end{equation*}
This objective reduces the amount of reports QoS Reporters have to send via network. The reason for this is that if a task or channel is part of more than one subgraph $G_i$, multiple QoS Managers require the measurement values of the element to be able to evaluate whether some of their constraints $constr(G_i)$ are violated.
\end{enumerate}

For some runtime graphs, objectives (1) and (2) are contradictory. Since we deem the network traffic caused by the QoS Reporters to be negligible, we believe condition (1) should be the primary focus.
Every allocation that optimizes the above objectives must however fulfill the following side conditions:
\begin{itemize}
 \item Every constraints lies within exactly one subgraph $G_i$ and is thus attended to by exactly one QoS Manager. Given that $constr(G_i)$ is the subset of runtime constraints whose sequence elements (tasks and channels) are included in $G_i$, the subgraphs must be chosen so that 
 \begin{equation*}
\bigcup_{0 \leq i < m}{constr(G_i)} = C
\end{equation*}
and all $constr(G_i)$ are pairwise disjoint. 
\item The subgraphs $G_i=(V_i, E_i)$ are of minimal size and thus do not contain any vertices irrelevant for the constraints. Given that $vertices(C)$ is the set of vertices contained in the sequences of $C's$ constraints, the following equation must hold:
\begin{equation*}
vertices(constr(G_i))=V_i
\end{equation*}
\end{itemize}
}

\paragraph{QoS Manager Setup}
After worker nodes have been allocated for all tasks, the master node will compute the subgraphs $G_i=(V_i,E_i)$ and send each one to a worker node so that it can start the QoS Manager background process.

\refAlg{ComputeQoSSetup} presents an overview of our approach to compute the subgraphs $G_i$. The algorithm is passed the user-defined job graph and job constraints and computes a set of QoS Manager allocations in the form of tuples $(w_i, G_i)$, where $w_i$ is the worker node supposed to run the QoS Manager for the (runtime) subgraph $G_i$. First, $GetConstrainedPaths()$  enumerates all paths (tuples of job vertices) through the job graph \dw{which} are covered by a job constraint. We do not provide pseudo-code for $GetConstrainedPaths()$ as the paths can be enumerated by simple depth-first traversal of the job graph. For each such path, we invoke $GetQoSManagers()$ to compute a set of $(w_i, G_i)$ tuples which is then merged into the set of already existing set of of QoS Manager allocations.

\begin{algorithm}
\caption{ComputeQoSSetup($JG, JC$)}
\labelAlg{ComputeQoSSetup}
\begin{algorithmic}[1]
\REQUIRE Job graph $JG$ and set of job constraints $JC$
\STATE $managers \leftarrow \emptyset$
\FORALL{$path \mbox{ \textbf{in} } GetConstrainedPaths(JG, JC)$}
\FORALL{$(w_i,G_i) \mbox{ \textbf{in} } GetQoSManagers(path)$}
\IF{$\exists(w_i, G^*_i) \in managers$}
\STATE $G^*_i \leftarrow mergeGraphs(G^*_i, G_i)$
\ELSE
\STATE $managers \leftarrow managers \cup \{(w_i,G_i)\}$
\ENDIF
\ENDFOR
\ENDFOR
\RETURN $managers$
\end{algorithmic}
\end{algorithm}

\refAlg{GetQoSManagers} computes the set of tuples $(w_i, G_i)$ that models which worker node runs a QoS Manager for the (runtime) subgraph $G_i$, where each $G_i$ is derived by splitting up the runtime graph corresponding to the given job graph path. First\dw{,} it uses $GetAnchorVertex()$ to determine an \textit{anchor job vertex} on the path. The anchor vertex serves as a starting point when determining the \ac{QoS} Managers and their subgraphs. \dw{The function} $PartitionByWorker()$ is used to split the anchor vertex into disjoint sets of runtime vertices that have been allocated to run on the same worker node. Using $GraphExpand()$ each such set $V_i$ of runtime vertices is then expanded to a runtime subgraph. This is done by traversing the runtime graph both forward\dw{s} and backwards (i.e.~with and against the edge direction of the \ac{DAG}), starting from the set of runtime vertices $V_i$.

\begin{algorithm}
\caption{GetQoSManagers($path$)}
\labelAlg{GetQoSManagers}
\begin{algorithmic}[1]
\REQUIRE $path \in JV^n$
\STATE $anchor \leftarrow GetAnchorVertex(path)$
\STATE $ret \leftarrow \emptyset$
\FORALL{$V_i \mbox{ \textbf{in} } PartitionByWorker(anchor)$}
\STATE $ret \leftarrow ret \cup \{(worker(V_i[0]), GraphExpand(V_i))\}$
\ENDFOR
\RETURN $ret$
\end{algorithmic}
\end{algorithm}

Finally, \refAlg{GetAnchorVertex} illustrates a simple heuristic to pick an anchor vertex for a constrained path through the job graph. The heuristic considers those job vertices as anchor candidates that have the highest worker count. It then picks the anchor candidate that has the job edge with the lowest number of runtime edges. To do so, $cntChan(jv, path)$ returns the number of runtime edges of the ingoing or outgoing job edge of $jv$ within the given path with the lowest number of runtime edges. The reasoning behind this is that anchor vertices with low numbers of runtime edges are more likely to produce smaller subgraphs for the QoS Managers when invoking $GraphExpand()$ in \refAlg{GetQoSManagers}.

\begin{algorithm}
\caption{GetAnchorVertex($path$)}
\labelAlg{GetAnchorVertex}
\begin{algorithmic}[1]
\REQUIRE $path=(jv_1, \ldots, jv_n) \in JV^n$
\STATE $ret \leftarrow \{jv_1, \ldots, jv_n\}$
\STATE $maxWork \leftarrow max\{cntWorkers(jv)| jv \in ret\}$
\STATE $ret \leftarrow ret \setminus \{jv \in ret| cntWorkers(jv) < maxWork\}$
\STATE $minEdge \leftarrow min\{cntEdge(jv, path)| jv \in ret\}$
\STATE $ret \leftarrow ret \setminus \{jv \in ret| cntEdge(jv,path) > minEdge\}$
\RETURN $ret[0]$
\end{algorithmic}
\end{algorithm}

\paragraph{QoS Reporter Setup}
The setup of the QoS Reporter processes is directly based on the QoS Manager setup. For each constrained runtime vertex $v \in V$ there is at least one QoS Manager with a subgraph $G_i=(V_i,E_i)$ and $v \in V_i$. The master node tracks this accordingly and instructs the QoS \dw{R}eporter to send measurement values of the running task to all interested QoS Managers. Channels are tracked in an analogous way.

\subsection{Reacting to Latency Constraint Violations}\labelSec{constraintViolationReaction}
Based on the workflow latency measured as described in \refSec{measuring}, each QoS Manager process can identify those sequences of its assigned runtime subgraph $G_i$ that violate their constraint and initiate countermeasures to improve latency. It will apply countermeasures until the constraint has been met or the necessary preconditions for applying countermeasures are not met anymore. In this case it will report the failed optimization attempt to the master node which in turn notifies the user who has to either change the job or revise the constraints.

Given a runtime subgraph $G_i=(V_i, E_i)$, a runtime sequence $S=(s_1, \ldots, s_n)$, and a violated latency constraint $(S, l, t)$, the QoS Manager attempts to eliminate the effect of improperly sized output buffers by adjusting the buffer sizes for each channel in $S$ individually and can apply dynamic task chaining to reduce latencies further. Buffer size adjustment is an iterative process which may increase or decrease buffer sizes at multiple channels, depending on the measured latencies. Note that after each run of the buffer adjustment procedure the QoS Manager waits until all latency measurement values based on the old buffer sizes have been flushed out. The conditions and procedures for changing buffer sizes and dynamic task chaining are outlined in the following sections.

\subsubsection{Adaptive Output Buffer Sizing}
\bl{For each channel $e \in E_i$ in the given sequence $S$ the QoS Manager permanently receives output buffer lifetime measurements (see \refSec{measuring}) and maintains a running average $oblt(e,t)$ of all measurements fresher than $t$ time units. It then estimates the \textit{average output buffer latency} of the data items that have passed through the channel during the last $t$ time units as $obl(e,t) = \frac{oblt(e,t)}{2}$. If $obl(e,t)$ supersedes both a sensible minimum threshold (for example $5$ \ac{ms}) and the task latency of the channel's source task, the QoS Manager sets the new output buffer size $obs^*(e)$  to}

\begin{equation}
obs^*(e) = max(\epsilon, obs(e)\times r^{obl(e,t)})
\end{equation}

where $\epsilon > 0$ is an absolute lower limit on the buffer size, $obs(e)$ is the current output buffer size, and $0<r<1$. We chose $r=0.98$ and $\epsilon = 200$ bytes as a default. This approach might reduce the output buffer size so much that most records do not fit inside the output buffer anymore, which is detrimental to both throughput and latency. Hence, if $obl(e)\approx 0$, we will increase the output buffer size to

\begin{equation}
obs^*(e) = min(\omega, s \times obs(e))
\end{equation} 

where $\omega >0$ is an upper bound for the buffer size and $s>1$. For our prototype we chose $s=1.1$.

Note that some channels may be in the subgraph of multiple  QoS Managers and that these may try to change its output buffer size at the same time. To deal with this, the worker node applies the buffer size update it receives first and discards any older updates. Additionally it will notify all relevant QoS Managers of the buffer size update with the next measurement value report so that they can keep their data up-to-date.

\subsubsection{Dynamic Task Chaining}
Task chaining pulls certain tasks into the same thread, thus eliminating the need for queues and thread-safe data item hand-over  between these tasks. In order to be able to chain a series of tasks $v_1, \ldots, v_n \in V_i$ within the constrained sequence S they need to fulfill the following conditions:
\begin{itemize}
 \item They all run as separate threads \dw{within} the same process on the worker node, which excludes any already chained tasks.
 \item The sum of the CPU utilizations of the task threads is lower than the capacity of one CPU core or a fraction thereof, for example $90$\% of a core. How such profiling information can be obtained has been described in \cite{battre.2010.mtags}. 
 \item They form a path through the QoS Manager's runtime subgraph, i.e.\ each pair $v_i, v_{i+1} \in V_i$ is connected by a channel $e=(v_i, v_{i+1}) \in E_i$.
 \item None of the tasks has more than one incoming and more than one outgoing channel, with the exception of the first task $v_1$ which is allowed to have multiple incoming channels and the last task $v_n$ which is allowed to have multiple outgoing channels.
\end{itemize}
The QoS Manager looks for the longest chainable series of tasks within the sequence. If it finds one, it instructs the worker node to chain the respective tasks. When chaining a series of tasks the worker node needs to take care of the input queues between them. There are two principal ways of doing this. The first one is to simply drop the existing input queues between these tasks. Whether this is acceptable or not depends on the nature of the workflow, for example in a video stream scenario it is usually acceptable to drop some frames. The second one is to halt the first task $v_1$ in the series and wait until the input queues between all of the subsequent tasks $v_2, \ldots, v_n$ in the chain have been drained. This will temporarily increase the latency in this part of the graph due to a growing input queue of $v_1$ that needs to be reduced after the chain has been established.

\subsection{Relation to Fault Tolerance}

In large clusters, individual nodes are likely to fail~\cite{mapreduce}. Therefore, it is important to point out how our proposed techniques to trade off high throughput against low latency at runtime affect the fault tolerance capabilities of current \dw{data processing frameworks}.

As these \dw{parallel} data processors mostly execute arbitrary black-box user code, currently the predominant approach to guard against execution failures is referred to as log-based rollback-recovery in literature~\cite{survey-rollback}. Besides sending the output buffers with the individual data items from the producing to the consuming task, the parallel processing frameworks additionally materialize these output buffers to a (distributed) file system. As a result, if a task or an entire worker node crashes, the data can be re-read from the file system and fed back into the re-started tasks. The fault tolerance in Nephele is also realized that way.

Our two proposed optimizations affect this type of fault tolerance mechanism in different ways: Our first approach, the adaptive output buffer sizing, is completely transparent to a possible data materialization because it does not change the framework's internal processing chain for output buffers but simply the size of these buffers. Therefore, if the parallel processing framework wrote output buffers to disk before the application of our optimization, it will continue to do so even if adaptive output buffer sizing is in operation.

For our second optimization, the dynamic task chaining, the situation is different. With dynamic task chaining activated, the data items passed from one task to the other no longer flow through the framework's internal processing chain. Instead, the task chaining deliberately bypasses this processing chain to avoid serialization/deserialization overhead and reduce latency. Possible materialization points may therefore be incomplete and useless for a recovery.

We addressed this problem by introducing an additional annotation to the Nephele job description. \dw{This annotation} prevents our system from applying dynamic task chaining \dw{to} particular parts of the \ac{DAG}. This way our streaming extension might lose one option to respond to violations of a provided latency goal, however, we are able to guarantee that Nephele's fault tolerance capabilities remain fully intact.

\section{Evaluation} \labelSec{evaluation}

\dw{After having} presented both the adaptive output buffer sizing and the dynamic task chaining for Nephele, we will now evaluate their impact based on an example job. To put the measured data into perspective, we also implemented the example job for another parallel data processing framework with streaming capabilities, namely Hadoop Online~\cite{hadooponline}.

We chose Hadoop Online as a baseline for comparison for three reasons: First, Hadoop Online is open-source software and was thus available for evaluation. Second, among all large-scale data processing frameworks with streaming capabilities, we think Hadoop Online currently enjoys the most popularity in the scientific community, which also makes it an interesting subject for comparison. Finally, in their research paper, the authors describe the continuous query feature of their system to allow for near-real-time analysis of data streams~\cite{mapreduceonline}. However, they do not provide any numbers on the actually achievable processing latency. Our experiments therefore also shed light on this question.

Please note that the experimental results presented in the following supersede the results from our previous publication~\cite{lohrmann.2012.hpdc}. Although the example job is nearly identical to the one used in the original paper, we were able to run the job on a significantly larger testbed ($200$ servers compared to ten servers) for this article. For the sake of a clearer presentation, we decided not to include the description of the original testbed and the experimental results again, however, would like to refer the interested reader to~\cite{lohrmann.2012.hpdc}.

\subsection{Job Description}

The job we use for the evaluation is motivated by the ``citizen journalism'' use case described in the introduction. We consider a web platform which offers its users to broadcast incoming video streams to a larger audience. However, instead of simple video transcoding which is done by existing video streaming platforms, our system additionally groups related video streams, merges them to a single stream, and \dw{also} augments the stream with additional information, such as Twitter feeds or other social network content. The idea is to provide the audience of the merged stream with a broader view of a situation by automatically aggregating related information from various sources.

In the following we will describe the structure of the job, first for Nephele and afterwards for Hadoop Online.

\subsubsection{Structure of the Nephele Job}\labelSec{nepheleJob}

\refFig{evaluationJob} depicts the structure of the Nephele evaluation job. The job consists of six distinct types of tasks. Each type of task is executed with a degree of parallelism of $m$, spread evenly across $n$ worker nodes.

\begin{figure}[t]
	\centering
		\includegraphics[width=7.5cm]{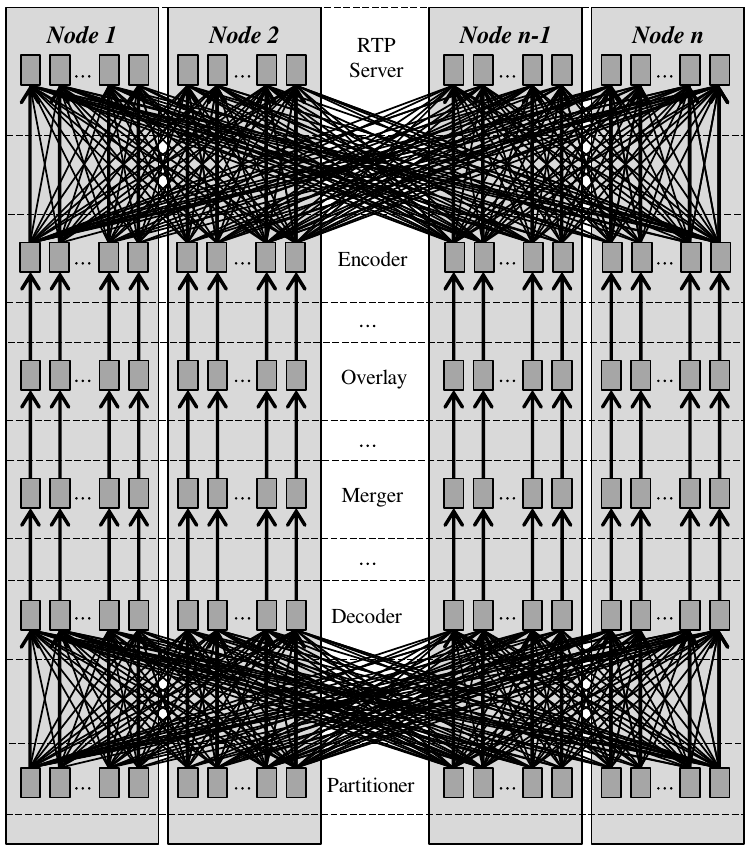}
		\caption{Runtime graph of the Nephele job}
	\labelFig{evaluationJob}
\end{figure}

The first tasks are of type \textit{Partitioner}. Each \textit{Partitioner} task acts as a TCP/IP server for incoming video feeds, receives H.264 encoded video streams, assigns them to a group of streams and forwards the video stream data to the \textit{Decoder} task responsible for streams of the assigned group. In the context of this evaluation job, we group video streams by a simple attribute which we expect to be attached to the stream as meta data, such as GPS coordinates. More sophisticated approaches to detect video stream correlations are possible but beyond the scope of our evaluation.

The \textit{Decoder} tasks are in charge of decompressing the encoded video packets into distinct frames which can then be manipulated later in the workflow. For the decoding process, we rely on the xuggle library~\cite{xuggle}.

Following the \textit{Decoder}, the next type of tasks in the processing pipeline are the \textit{Merger} tasks. \textit{Merger} tasks consume frames from grouped video streams and merge the respective set of frames to a single output frame. In our implementation the merge step simply consists of tiling the individual input frames in the output frame. 

After having merged the grouped input frames, the \textit{Merger} tasks send their output frames to the next task type in the pipeline, the \textit{Overlay} tasks. An \textit{Overlay} task augments the merged frames with information from additional related sources. For the evaluation, we designed each \textit{Overlay} task to draw a marquee of Twitter feeds inside the video stream, which are picked based on locations close to the GPS coordinates attached to the video stream.

The output frames of the \textit{Overlay} tasks are encoded back into the H.264 format by a set of \textit{Encoder} tasks and then passed on to tasks of type \textit{RTP Server}. These tasks represent the sink of the streams in our workflow. Each task of this type passes the incoming video streams on to an RTP server which then offers the video to an interested audience.

\subsubsection{Structure of the Hadoop Online Job}\labelSec{hadoopJob}

For Hadoop Online, the example job exhibits a similar structure as for Nephele, however, the six distinct tasks have been distributed among the map and reduce functions of two individual MapReduce jobs. During the experiments on Hadoop Online, we executed the exact same task code as for Nephele apart from some additional wrapper classes we had to write in order to achieve interface compatibility.

\begin{figure}[t]
	\centering
		\includegraphics[width=7.5cm]{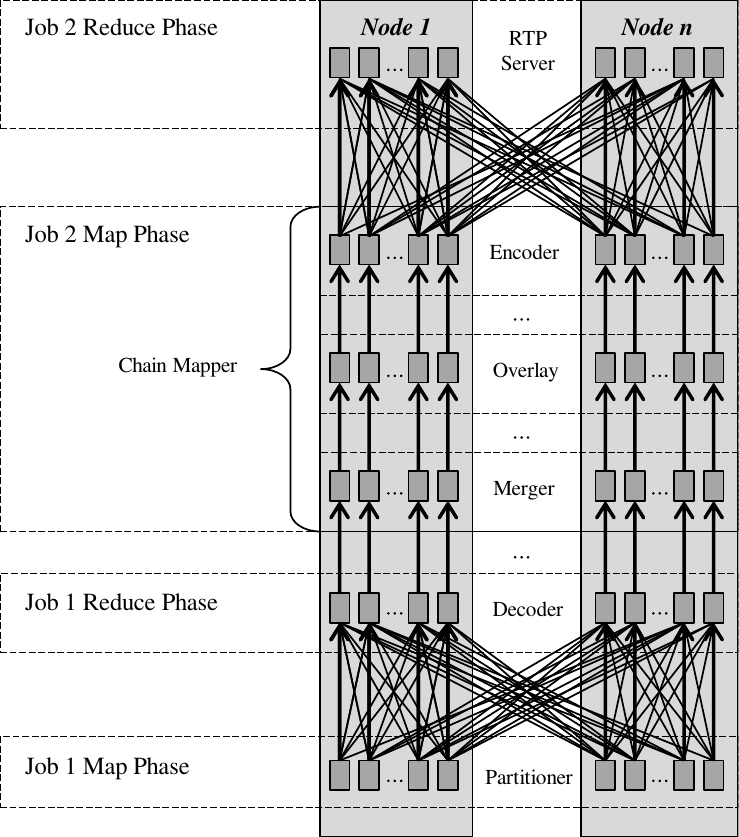}
		\caption{Runtime graph of the Hadoop Online job}
	\labelFig{evaluationJobHop}
\end{figure}

As illustrated in~\refFig{evaluationJobHop} we inserted the initial \textit{Partitioner} task into the map function of the first MapReduce job. Following the continuous query example from the Hadoop Online website, the task basically ``hijacks'' the map slot with an infinite loop and waits for incoming H.264 encoded video streams. Upon the reception of the stream packet, the packet is put out with a new key, such that all video streams within the same group will arrive at the same parallel instance of the reducer. The reducer function then accommodates the previously described \textit{Decoder} task. As in the Nephele job, the \textit{Decoder} task decompresses the encoded video packets into individual frames.

The second MapReduce job starts with the three tasks \textit{Merger}, \textit{Overlay}, and \textit{Encoder} in the map phase. Following our experiences with the computational complexity of these tasks from our initial Nephele experiments, we decided to use a Hadoop chain mapper and execute all of these three tasks consecutively within a single map process. Finally, in the reduce phase of the second MapReduce job, we placed the task \textit{RTP Server}. The \textit{RTP Server} tasks again represented the sink of our data streams.

In comparison to the classic Hadoop, the evaluation job exploits two distinct features of the Hadoop Online prototype, i.e. the support for continuous queries and the ability to express dependencies between different MapReduce jobs. The continuous query feature allows to stream data from the mapper directly to the reducer. The reducer then runs a moving window over the received data. We set the window size to $100$ \ac{ms} during the experiments. For smaller window sizes, we experienced no significant effect on the latency.

 \begin{figure}[t]
	\centering
		\includegraphics{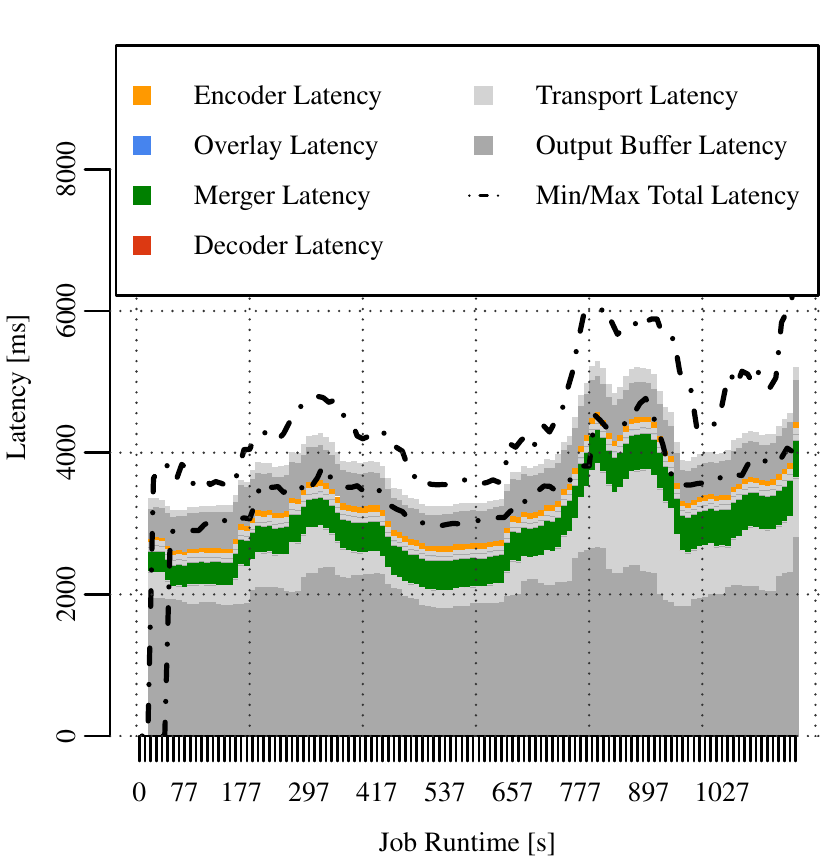}
		\caption{Latency w/o optimizations ($6400$ video streams, degree of parallelism $m=800$, $32$ KB fixed output buffer size)}
	\labelFig{noOptimization}
 \end{figure}

\subsection{Experimental Setup}

We executed our evaluation job on a cluster of $n=200$ commodity servers. Each server was equipped with an Intel Xeon E3-1230 V2 $3.3$ GHz (four real CPU cores plus hyper-threading activated) and $16$ GB RAM. The nodes were connected via regular Gigabit Ethernet links and ran Linux (kernel version 3.3.8) as well as and Java 1.6.0.26, which is required by Nephele's worker component. Additionally, each \dw{server} launched a \ac{NTP} daemon to maintain clock synchronization among the workers. During the entire experiment, the measured clock skew was below $2$ \ac{ms} among the machines.

Each \dw{of the worker nodes} ran eight tasks of type \textit{Decoder}, \textit{Merger}, \textit{Overlay} and \textit{RTP Server}, respectively. The number of incoming video streams was fixed for each experiment and they were evenly distributed over the \textit{Partitioner} tasks. We always grouped and subsequently merged four streams into one aggregated video stream. Each video stream had a resolution of $320 \times 240$ pixels and was H.264 encoded. The initial output buffer size was $32$ KB. Unless noted otherwise, all tasks had a degree of parallelism of \dw{$m=800$}.

Those experiments that were conducted on Nephele with latency constraints in place, specified one runtime constraint $c=(S,l,t)$ for each possible runtime sequence

\begin{equation}
S=(e_1, v_D, e_2, v_M, e_3, v_O, e_4, v_E, e_5) 
\end{equation}

where $v_D, v_M, v_O, v_E$ \dw{represent tasks of the types} \textit{Decoder}, \textit{Merger}, \textit{Overlay} and \textit{Encoder}, respectively. The altogether $512\times10^6$ constraints specified the same upper latency bound $l=300$ \ac{ms} over the data items within the past $t=15$ seconds. The measurement interval on the worker nodes was set to $15$ seconds, too.

\begin{figure}[t]
 	\centering
 		\includegraphics{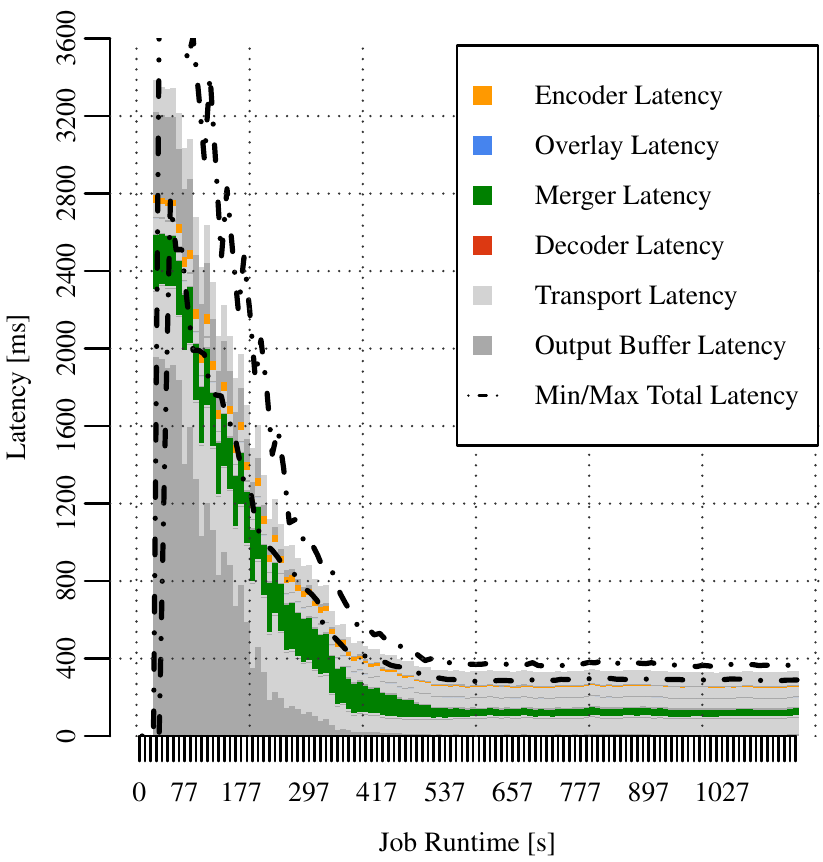}
 		\caption{Latency with adaptive buffer sizing ($6400$ video streams, degree of parallelism $m=800$, $32$ KB initial output buffer size)}
 	\labelFig{adaptiveBuffersizeOptimization}
 \end{figure}
 
  \begin{figure}[t]
 	\centering
 		\includegraphics{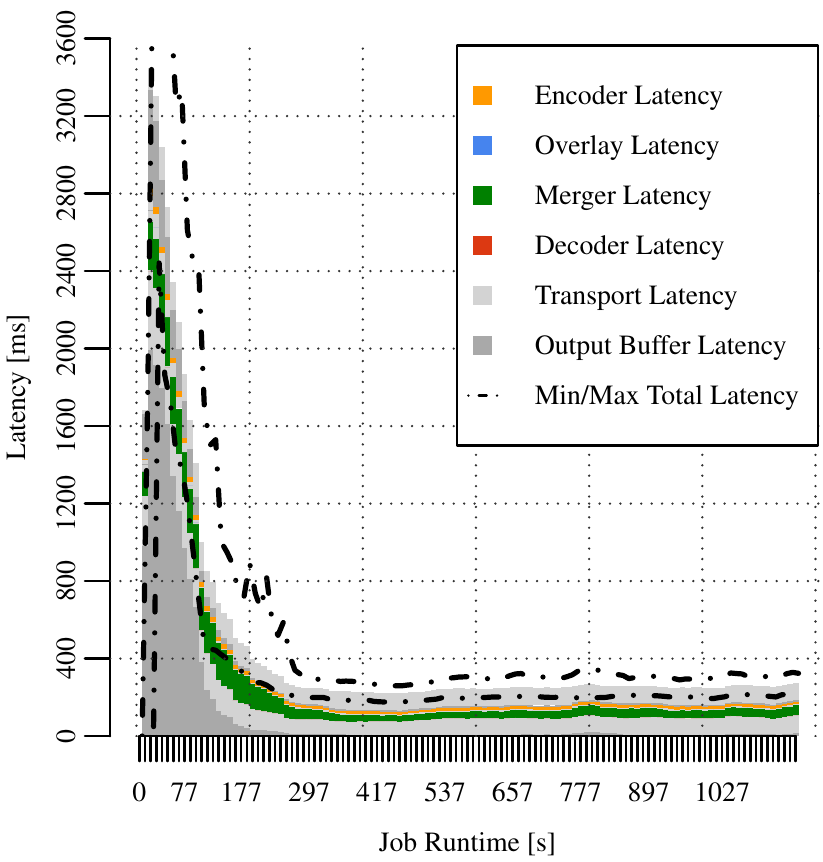}
 		\caption{Latency with adaptive buffer sizing and \dw{dynamic} task chaining ($6400$ video streams, degree of parallelism $m=800$, $32$ KB initial output buffer size)}
 	\labelFig{adaptiveBufferAndChainOptimization}
 \end{figure}

\subsection{Experimental Results}\labelSec{experimentalResults}
We evaluated our approach on the Nephele framework with the job described in \refSec{nepheleJob} in three scenarios which are (1) without any kind of latency optimizations (2) with adaptive output buffer sizing and (3) with adaptive output buffer sizing as well as dynamic task chaining. As a baseline for comparison with other frameworks we evaluated the Hadoop Online \dw{job} described in \refSec{hadoopJob} on the same testbed.

\subsubsection{Latency without Optimizations}\labelSec{woOpt}

First, we ran the Nephele job with constraints in place but prevented the QoS Managers from applying any optimizations. \refFig{noOptimization} summarizes the aggregated measurement data of all QoS Managers. As described in \refSec{measuring}, each QoS Manager maintains running averages of the measured latencies of its tasks and channels. Each sub-bar displays the arithmetic mean over the running averages for tasks/channels of the same type. For the plot, each channel latency is split up into mean output buffer latency (dark gray) and mean transport latency (light gray), which is the remainder of the channel latency after subtracting output buffer latency. Hence, the total height of each bar is the sum of the arithmetic means of all task/channel latencies and gives an impression of the current overall workflow latency. The dot-dashed lines provide information about the distribution of measured sequence latencies (min and max).

The total workflow latency fluctuated between $3.5$ and $5.5$ seconds. The figure clearly shows that output buffer and channel latencies massively dominated the total workflow latency, so much in fact that most task latencies are hardly visible at all. The main reason for this is the output buffer size of $32$ KB which was too large for the compressed video stream packets between \textit{Partitioner} and \textit{Decoder} tasks, as well as \textit{Encoder} and \textit{RTP Server} tasks. These buffers sometimes took longer than $1$ second to be filled and when they were placed into the input queue of a \textit{Decoder} they would take a while to \dw{be} processed. The situation was even worse between the \textit{Encoder} and \textit{RTP Server} tasks as the number of streams had been reduced by four and thus it took even longer to fill a $32$ KB buffer. Between the \textit{Decoder} and \textit{Encoder} tasks the channel latencies were much lower since the initial buffer size was a better fit for the decompressed images.

Another consequence of the buffer size were large variations in total workflow latency that stemmed from the fact that task threads such as the \textit{Decoder} could not fully utilize their CPU time because they fluctuated between idling due to input starvation and full CPU utilization once a buffer had arrived.

The anomalous task latency of the \textit{Merger} task is caused by the way we measure task latencies and limitations of our frame merging implementation. Frames that needed to be grouped always arrived in different buffers. With large buffers arriving at a slow rate the \textit{Merger} task did not always have images from all grouped streams available and would not produce any merged frames. This caused the framework to measure high task latencies (see \refSec{taskLatency}).

\subsubsection{Latency with Adaptive Output Buffer Sizing}\labelSec{wAdaptiveBufferSize}
\refFig{adaptiveBuffersizeOptimization} shows the results when using only adaptive buffer sizing to meet latency constraints. The structure of the plot is identical to \refFig{noOptimization}.

Our approach to adaptive buffer sizing quickly reduced the buffer sizes on the channels between \textit{Partitioner} and \textit{Decoder} tasks, as well as \textit{Encoder} and \textit{RTP server} tasks. The effect of this is clearly visible in the diagram, with an initial workflow latency of $3.4$ seconds that is reduced to $340$ \ac{ms} on average and $380 $ \ac{ms} in the worst case. The latency constraint of $300$ \ac{ms} has not been met, however we attained a latency improvement of one order of magnitude compared to the unoptimized Nephele job.

The convergence phase at the beginning of the job during which buffer sizes were decreased took approx.\ $9$ minutes. There are several reasons for this phenomenon. First, as the workers started with output buffers whose lifetime was sometimes larger than the measurement interval there often was not enough measurement data for the QoS Managers to act upon during this phase. In this case it waited until enough measurement data were available before checking for constraint violations. Second, after each output buffer size change a QoS Manager waits until all old measurements for the respective channel have been flushed out before revisiting the violated constraint, which took at least $15$ seconds each time.

\subsubsection{Latency with Adaptive Output Buffer Sizing and \dw{Dynamic} Task Chaining}\labelSec{wAdaptiveBufferSizeAndChaining}

\refFig{adaptiveBufferAndChainOptimization} shows the results when using adaptive buffer sizing and dynamic task chaining. The latency constraints were identical to those in \refSec{wAdaptiveBufferSize} and the structure of the plot is again identical to \refFig{noOptimization}.

Our task chaining approach chose to chain the \textit{Decoder}, \textit{Merger}, \textit{Overlay} and \textit{Encoder} tasks because the sum of their CPU utilizations did not fully saturate one CPU core.

After the initial calibration phase, the total workflow latency stabilized at an average of around $270$ \ac{ms} and a maximum of approx.\ $320$ \ac{ms}. This finally met all defined latency constraints, which caused the QoS Managers to not trigger any further actions. In our case this constituted another $26$\% improvement in latency compared to not using \dw{dynamic} task chaining and an improvement by a factor of at least $13$ compared to the unoptimized Nephele job.

\begin{figure}[t]
 	\centering
 		\includegraphics{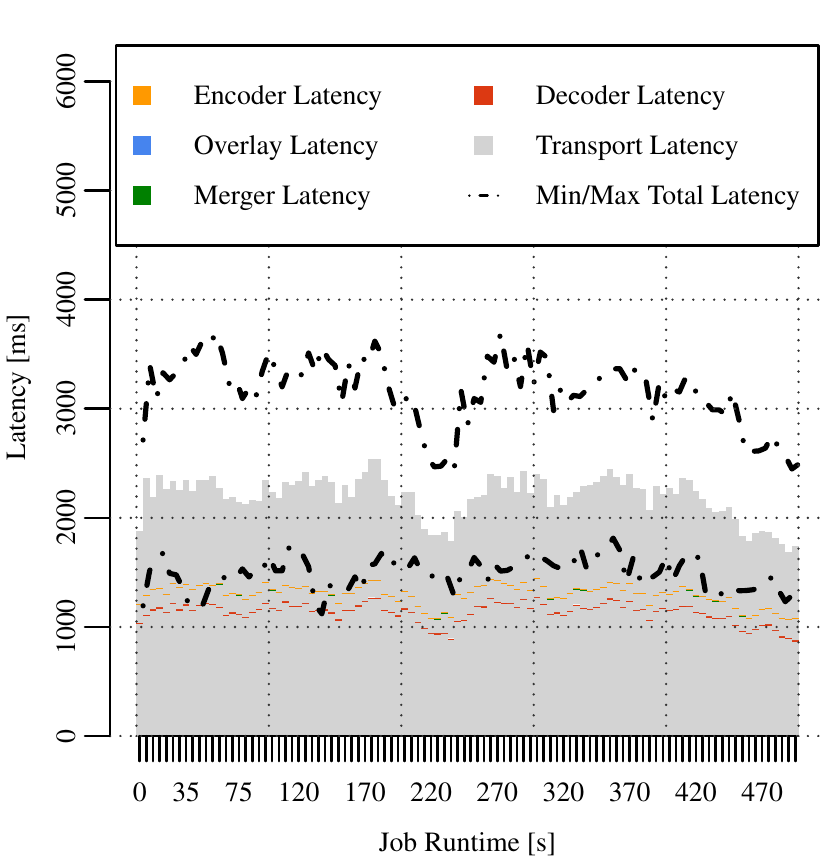}
 		\caption{Latency in Hadoop Online ($80$ video streams, degree of parallelism $m=10$, $100$ \ac{ms} window size)}
 	\labelFig{hopLatencyPlot}
\end{figure}

\subsubsection{Latency in Hadoop Online}\labelSec{hopEvaluation}

\refFig{hopLatencyPlot} shows a bar plot of the task and channel latencies obtained from the experiments with the Hadoop Online prototype. The plot's structure is again identical to \refFig{noOptimization}, however the output buffer latency has been omitted as these measurements are not offered by Hadoop Online.

Similar to the unoptimized Nephele job, the overall processing latency of Hadoop Online was clearly dominated by the channel latencies. Except for the tasks in the chain mapper, each data item experienced an average latency of up to one second when being passed on from one task to the next.

Due to technical difficulties with the Hadoop Online prototype, we were forced to reduce the degree of parallelism for the experiment to $m=10$ with only one deployed processing pipeline per host. The number of incoming streams was reduced to $80$ in order to match the relative workload (eight streams per pipeline) of the previous Nephele experiments. A positive effect of this reduction is a significantly lower task latency of the \text{Merger} task because, with fewer streams, the task had to wait less often for an entire frame group to be completed.

Apart from the size of the window reducer, we also varied the number of worker nodes $n$ in the range of $2$ to $10$ as a side experiment. However, we did not observe a significant effect on the channel latency either.

\section{Related Work} \labelSec{relatedWork}

Over the past decade stream processing has been the subject of vivid research. \dw{With regard of their scalability}, the existing approaches \dw{can essentially be subdivided into} three categories: Centralized, distributed, and massively-parallel stream processors.

\dw{Initially,} several centralized systems for stream processing have been proposed, such as Aurora~\cite{aurora} and STREAM~\cite{stream01,stream03}. Aurora is a DBMS for continuous queries that are constructed by connecting a set of predefined operators to a \ac{DAG}. The stream processing engine schedules the execution of the operators and uses load shedding, i.e.\ dropping intermediate tuples to meet \ac{QoS} goals. At the end points of the graph, user-defined \ac{QoS} functions are used to specify the desired latency and which tuples can be dropped. STREAM presents additional strategies for applying load-shedding, such as probabilistic exclusion of tuples. While these systems have useful properties such as respecting latency requirements, they run on a single host and do not scale well with rising data rates and numbers of data sources.

Later systems such as Aurora*/Medusa~\cite{aurorastarmedusa} support distributed processing of data streams. An Aurora* system is a set of Aurora nodes that cooperate via an overlay network within the same administrative domain. In Aurora* the nodes can freely relocate load by decentralized, pairwise  exchange of Aurora stream operators. Medusa integrates many participants such as several sites running Aurora* systems from different administrative domains into a single federated system. Borealis~\cite{borealis} extends Aurora*/Medusa and introduces, amongst other features, a refined \ac{QoS} optimization model where the effects of load shedding on \ac{QoS} can be computed at every point in the data flow. This enables the optimizer to find better strategies for load shedding.

The third category of possible stream processing systems is constituted by massively-parallel data processing systems. In contrast to the previous two categories, these systems have been designed to run on hundreds or even thousands of nodes in the first place and to efficiently transfer large data volumes between them. Traditionally, those systems have been used to process finite blocks of data stored on distributed file systems. However, many of the newer systems like Dryad~\cite{dryad}, Hyracks~\cite{hyracks}, CIEL~\cite{ciel}, or our Nephele framework~\cite{nephelejournal} allow to assemble complex parallel data flow graphs and to construct pipelines between the individual parts of the flow. Therefore, these parallel data flow systems in general are also suitable for streaming applications.

Recently, a series of systems have been introduced which aim to carry over the popular MapReduce programming model to parallel stream processing.

The first \dw{work} in this space was arguably Hadoop \dw{O}nline, described in~\cite{mapreduceonline}. As already mentioned in~\refSec{hadoopJob} the developers of Hadoop \dw{O}nline extended the original Hadoop system by the ability to stream intermediate results from the map to the reduce tasks as well as the possibility to pipeline data across different MapReduce jobs. To facilitate these new features, they extended the semantics of the classic reduce function by time-based sliding windows. Li et al.~\cite{li2011} picked up this idea and further improved the suitability of Hadoop-based systems for continuous streams by replacing the sort-merge implementation for partitioning by a new hash-based technique.

The Muppet system~\cite{muppet} also focuses on the parallel processing of continuous stream data while preserving a MapReduce-like programming abstraction. However, the authors decided to replace the reduce function by a more generic update function to allow for greater flexibility when processing intermediate data with identical keys. Muppet also aims to support near-real-time processing latencies. Unfortunately, the paper provides only few details on how data is actually passed between tasks (and hosts). We assume however that the system uses a communication scheme unlike the one we explained in~\refSec{principalFrameworkProperties}.

The systems S4~\cite{s4} and Storm~\cite{storm} can also be classified as massively-parallel data processing systems with a clear emphasis on low latency. Their programming abstraction is not MapReduce but allows developers to assemble arbitrarily complex \acs{DAG} of processing tasks. Similar to Muppet, both systems do not necessarily follow the design principles explained  in~\refSec{principalFrameworkProperties}. For example, Twitter Storm does not use intermediate queues to pass data items from one task to the other. Instead, data items are passed directly between tasks using batch messages on the network level to achieve a good balance between latency and throughput.

None of the systems from the third category has so far offered the capability to express high-level \ac{QoS} goals as part of the job description and let the system optimize towards these goals independently, as it was common for previous systems from category one and two.

\section{Conclusion and Future Work} \labelSec{conclusion}

\dw{The growing number of commodity devices capable of producing continuous data streams promises to unlock a whole new class of interesting and innovative use cases, however also raises concerns with regard to the scalability of existing stream processors. While the individual data streams may be characterized by comparably low data volumes, processing them at scale can quickly call for large compute clusters and platforms for data-intensive computing.

In this paper, we therefore examined the suitability of existing massively-parallel data processing frameworks for large-scale stream processing. We identified common design principles among those frameworks and highlighted two new techniques, \textit{adaptive output buffer sizing} and \textit{dynamic task chaining}, which allow them to dynamically trade off higher throughput against lower processing latency. Based on our parallel data processor Nephele, we thereupon proposed a highly distributed scheme to detect violations of user-defined \ac{QoS} constraints at runtime and illustrated how both of our techniques can help to automatically mitigate those. Through a sample video streaming use case on a large-scale cluster system, we found that our strategies can improve workflow latency by a factor of at least $13$ while preserving the required data throughput.}

We see the need for future work on this topic in several areas. The Nephele framework is part of a bigger software stack for massively-parallel data analysis developed within the Stratosphere project~\cite{stratosphere}. Therefore, extending the streaming capabilities to the upper layers of the stack, in particular to the PACT programming model~\cite{nephelepacts}, is of future interest. Furthermore, we plan to explore strategies for other \ac{QoS} goals such as jitter and throughput that exploit the capability of a cloud to elastically scale on demand.

\dw{In general we think our work marks an important first step towards introducing \ac{QoS} considerations in the domain of massively-parallel data processing and helps to support new classes of \ac{QoS}-sensitive streaming applications at scale.}

\bibliographystyle{spmpsci}      
\bibliography{ccjournal12}   

\begin{thebibliography}{10}
\providecommand{\url}[1]{{#1}}
\providecommand{\urlprefix}{URL }
\expandafter\ifx\csname urlstyle\endcsname\relax
  \providecommand{\doi}[1]{DOI~\discretionary{}{}{}#1}\else
  \providecommand{\doi}{DOI~\discretionary{}{}{}\begingroup
  \urlstyle{rm}\Url}\fi

\bibitem{hadooponline}
{Hadoop Online Prototype - Google Project Hosting}.
\newblock \url{http://code.google.com/p/hop/} (2012)

\bibitem{justin.tv}
{Justin.tv - Streaming live video broadcasts for everyone}.
\newblock \url{http://www.justin.tv/} (2012)

\bibitem{livestream}
{Livestream - Be There}.
\newblock \url{http://www.livestream.com/} (2012)

\bibitem{storm}
{nathanmarz/storm - GitHub}.
\newblock \url{https://github.com/nathanmarz/storm} (2012)

\bibitem{stratosphere}
{Stratosphere - Above the Clouds}.
\newblock \url{http://stratosphere.eu/} (2012)

\bibitem{ustream}
{USTREAM, You're On.}
\newblock \url{http://www.ustream.tv/} (2012)

\bibitem{hadoop}
{Welcome to Apache Hadoop!}
\newblock \url{http://http://hadoop.apache.org/} (2012)

\bibitem{xuggle}
{Xuggle}.
\newblock \url{http://http://www.xuggle.com/} (2012)

\bibitem{borealis}
Abadi, D., Ahmad, Y., Balazinska, M., Cetintemel, U., Cherniack, M., Hwang, J.,
  Lindner, W., Maskey, A., Rasin, A., Ryvkina, E., et~al.: The design of the
  {Borealis} stream processing engine.
\newblock In: Second Biennial Conference on Innovative Data Systems Research,
  CIDR '05, pp. 277--289 (2005)

\bibitem{aurora}
Abadi, D., Carney, D., {\c{C}}etintemel, U., Cherniack, M., Convey, C., Lee,
  S., Stonebraker, M., Tatbul, N., Zdonik, S.: Aurora: A new model and
  architecture for data stream management.
\newblock The VLDB Journal \textbf{12}(2), 120--139 (2003)

\bibitem{aldinucci.1999.pdcs}
Aldinucci, M., Danelutto, M.: Stream parallel skeleton optimization.
\newblock In: Proc.\ of the 11th IASTED International Conference on Parallel
  and Distributed Computing and Systems, PDCS '99, pp. 955--962. IASTED/ACTA
  (1999).
\newblock
  \urlprefix\url{ftp://ftp.di.unipi.it/pub/Papers/aldinuc/302-114.ps.gz}

\bibitem{alexandrov.2011.btw}
Alexandrov, A., Ewen, S., Heimel, M., Hueske, F., Kao, O., Markl, V., Nijkamp,
  E., Warneke, D.: {MapReduce} and {PACT} - comparing data parallel programming
  models.
\newblock In: Proc.\ of the 14th Conference on Database Systems for Business,
  Technology, and Web, BTW '11, pp. 25--44. GI (2011)

\bibitem{stream01}
Babu, S., Widom, J.: Continuous queries over data streams.
\newblock SIGMOD Rec. \textbf{30}, 109--120 (2001)

\bibitem{nephelepacts}
Battr\'{e}, D., Ewen, S., Hueske, F., Kao, O., Markl, V., Warneke, D.:
  Nephele/{PACT}s: {A} programming model and execution framework for web-scale
  analytical processing.
\newblock In: Proc.\ of the 1st ACM symposium on Cloud computing, SoCC '10, pp.
  119--130. ACM (2010)

\bibitem{battre.2010.mtags}
Battr\'{e}, D., Hovestadt, M., Lohrmann, B., Stanik, A., Warneke, D.: Detecting
  bottlenecks in parallel {DAG}-based data flow programs.
\newblock In: Proc.\ of the 2010 IEEE Workshop on Many-Task Computing on Grids
  and Supercomputers, MTAGS '10, pp. 1--10. IEEE (2010)

\bibitem{hyracks}
Borkar, V., Carey, M., Grover, R., Onose, N., Vernica, R.: Hyracks: A flexible
  and extensible foundation for data-intensive computing.
\newblock In: Proc.\ of the 2011 IEEE 27th International Conference on Data
  Engineering, ICDE '11, pp. 1151--1162. IEEE (2011).
\newblock \doi{http://dx.doi.org/10.1109/ICDE.2011.5767921}.
\newblock \urlprefix\url{http://dx.doi.org/10.1109/ICDE.2011.5767921}

\bibitem{aurorastarmedusa}
Cherniack, M., Balakrishnan, H., Balazinska, M., Carney, D., Cetintemel, U.,
  Xing, Y., Zdonik, S.: Scalable distributed stream processing.
\newblock In: Proc.\ of the First Biennial Conference on Innovative Data
  Systems Research, CIDR '03, pp. 257--268 (2003)

\bibitem{mapreduceonline}
Condie, T., Conway, N., Alvaro, P., Hellerstein, J.M., Elmeleegy, K., Sears,
  R.: {MapReduce} {Online}.
\newblock In: Proc.\ of the 7th USENIX conference on Networked systems design
  and implementation, NSDI '10, pp. 21--21. USENIX Association (2010)

\bibitem{mapreduce}
Dean, J., Ghemawat, S.: {MapReduce}: {Simplified} data processing on large
  clusters.
\newblock Communications of the ACM \textbf{51}(1), 107--113 (2008)

\bibitem{survey-rollback}
Elnozahy, E.N.M., Alvisi, L., Wang, Y.M., Johnson, D.B.: A survey of
  rollback-recovery protocols in message-passing systems.
\newblock ACM Comput. Surv. \textbf{34}(3), 375--408 (2002).
\newblock \doi{http://doi.acm.org/10.1145/568522.568525}

\bibitem{dryad}
Isard, M., Budiu, M., Yu, Y., Birrell, A., Fetterly, D.: Dryad: {Distributed}
  data-parallel programs from sequential building blocks.
\newblock ACM SIGOPS Operating Systems Review \textbf{41}(3), 59--72 (2007)

\bibitem{muppet}
Lam, W., Liu, L., Prasad, S., Rajaraman, A., Vacheri, Z., Doan, A.: Muppet:
  Mapreduce-style processing of fast data.
\newblock Proc.\ VLDB Endow. \textbf{5}(12), 1814--1825 (2012)

\bibitem{li2011}
Li, B., Mazur, E., Diao, Y., McGregor, A., Shenoy, P.: A platform for scalable
  one-pass analytics using mapreduce.
\newblock In: Proceedings of the 2011 ACM SIGMOD International Conference on
  Management of data, SIGMOD '11, pp. 985--996. ACM, New York, NY, USA (2011)

\bibitem{lohrmann.2012.hpdc}
Lohrmann, B., Warneke, D., Kao, O.: Massively-parallel stream processing under
  {QoS} constraints with {N}ephele.
\newblock In: Proceedings of the 21st International Symposium on
  High-Performance Parallel and Distributed Computing, HPDC '12, pp. 271--282.
  ACM, New York, NY, USA (2012)

\bibitem{stream03}
Motwani, R., Widom, J., Arasu, A., Babcock, B., Babu, S., Datar, M., Manku, G.,
  Olston, C., Rosenstein, J., Varma, R.: Query processing, approximation, and
  resource management in a data stream management system.
\newblock In: First Biennial Conference on Innovative Data Systems Research,
  CIDR '03, pp. 245--256 (2003)

\bibitem{ciel}
Murray, D., Schwarzkopf, M., Smowton, C., Smith, S., Madhavapeddy, A., Hand,
  S.: {CIEL}: {A} universal execution engine for distributed data-flow
  computing.
\newblock In: Proc.\ of the 8th USENIX conference on Networked systems design
  and implementation, NSDI '11, pp. 9--9. USENIX Association (2011)

\bibitem{s4}
Neumeyer, L., Robbins, B., Nair, A., Kesari, A.: S4: Distributed stream
  computing platform.
\newblock In: 2010 IEEE International Conference on Data Mining Workshops,
  ICDMW '10, pp. 170--177. IEEE (2010)

\bibitem{nephelejournal}
Warneke, D., Kao, O.: Exploiting dynamic resource allocation for efficient
  parallel data processing in the cloud.
\newblock IEEE Transactions on Parallel and Distributed Systems \textbf{22}(6),
  985--997 (2011)

\end{thebibliography}

\end{document}